\documentclass[aps,pre,reprint,amsmath,amssymb]{revtex4-1}

\usepackage{graphicx}
\usepackage[retainorgcmds]{IEEEtrantools}
\usepackage{color}
\usepackage[retainorgcmds]{IEEEtrantools}
\usepackage{graphicx}
\usepackage{mathrsfs}
\usepackage{amsmath}
\usepackage{amssymb}
\usepackage{color}
\usepackage{amsfonts}

\begin{document}

\title{Thermal rectification in anharmonic chains under an energy-conserving noise}
\date{\today}
\author{Pedro H. Guimar\~aes$^{\,1}$, Gabriel T. Landi$^{\,2}$
and M\'ario J. de Oliveira$^{\,1}$}
\affiliation{$^1$Instituto de F\'isica, Universidade de S\~ao Paulo,
Caixa Postal 66318, 05314-970 S\~ao Paulo, Brazil,
$^2$Universidade Federal do ABC,  09210-580 Santo Andr\'e, Brazil}

\begin{abstract}

Systems in which the heat flux depends on the direction of the flow are said to present thermal rectification. 
This effect has attracted much theoretical and experimental interest in recent years. 
However, in most theoretical models the effect is found to vanish in the thermodynamic limit, in disagreement with experiment. 
The purpose of this paper is to show that the rectification may be restored by including an energy-conserving  noise which randomly flips the velocity of the particles with a certain rate $\lambda$.
It is shown that as long as $\lambda$ is non-zero, the rectification remains finite in the thermodynamic limit. 
This is illustrated in a classical harmonic chain subject to a quartic pinning potential (the $\Phi^4$ model) and coupled to heat baths by Langevin equations.

\end{abstract}

\maketitle{}

%----------------------------------------------------------
\section{Introduction}

When a system is placed between two reservoirs maintained at different temperatures, a heat current $J$ is established from the hotter reservoir to the colder. 
For most materials, this current is independent of the direction of the flow. 
But if the material is asymmetric, this may not be the case. 
When the heat current depends on the direction of the flow, we say the system presents \emph{thermal rectification} (in analogy with the electric rectification of diodes).
This topic has recently been the subject of intensive investigations 
\cite{Hu2006,Terraneo2002,Li2004a,Pereira2004,Li2005,
Wang2007,Lo2008,Pereira2008,Scheibner2008,Hu2009,Kobayashi2009,Yang2009,
Zhang2009,Pereira2010b,Pereira2010,Pereira2011,Pereira2011b,Shah2012,
Wang2012a,Pereira2013a,Thingna2013,Avila2013,Romero-Bastida2013,
Werlang2014,Kubytskyi2014,Nefzaoui2014,Pereira2011b,Landi2014b,Landi2015a}, 
motivated in part by the possibility of constructing devices operating exclusively with heat current (for reviews, see \cite{Roberts2011,Li2012}).

There are at least two key ingredients to the existence of thermal rectification. 
The first is an inherent asymmetry of the system which breaks the invariance   under bath  reversal.
The second is a temperature dependence of the thermal conductivity,  which is necessary to induce an overlap between the phonon spectra of the left and right parts of the system \cite{Li2004a,Pereira2011}.
This is the reason why the classical harmonic chain \cite{Rieder1967} presents no rectification, irrespective of any asymmetries present. 

In the thermodynamic limit, however, these two ingredients may  not  suffice to maintain a finite rectification. 
Indeed, most theoretical models and numerical simulations predict a rectification that  vanishes  in the thermodynamic limit; i.e., for a macroscopically large sample. 
This is in clear disagreement with experiment\cite{Starr1936,Brattain1951,Chang2006,Scheibner2008,Hu2009,Kobayashi2009,Yang2009,Nefzaoui2014}.

%This discrepancy  indicates that an additional ingredient is still necessary. 
%, where a finite rectification is observed even in macroscopically large systems 

The purpose of this paper is to show that this discrepancy may be resolved by including an additional energy-conserving noise.
This noise represents the weak interaction between the system and other degrees of freedom that are always present in real solids. 
It therefore serves as an ad-hoc implementation of the extraordinary complexity of real systems. 
In the past it was used to solve the problem of anomalous heat conduction \cite{Bolsterli1970,Bonetto2004,Bernardin2005,Dhar2011,Landi2014a,Landi2013a}: 
 The harmonic chain coupled to two baths at different temperatures has a ballistic heat flow \cite{Rieder1967}, but when this noise is included the flux becomes diffusive, as expected from Fourier's law.
The idea of an energy-conserving noise is also extensively used in quantum optics, where it is referred to as dephasing \cite{Gardiner2004}. 
In either case, the assumption that the noise is energy-conserving is justified when the coupling to the additional degrees of freedom is weak.

There are several ways to implement the energy-conserving noise. 
One is through self-consistent reservoirs  \cite{Bolsterli1970,Bonetto2004}: All particles in the system are coupled to auxiliary Langevin baths at different temperatures, which are adjusted so that in the steady-state no heat flows through them. 
The implementation which will be used here was first proposed in Ref.~\cite{Dhar2011} by Dhar, Venkateshan and Lebowitz, and was also studied by some of us in Refs.~\cite{Landi2014a,Landi2013a}. It is somewhat similar to the noise studied by Bernardin and Olla in Ref.~\cite{Bernardin2005}. 
The idea is that the particles in the system may suffer elastic collisions with other degrees of freedom, whose effect is to randomly rotate the velocity vector of each particle by a small (random) angle. Since the kinetic energy depends only on the magnitude of the velocity, this conserves the kinetic energy and therefore the total energy of the system. 

In one-dimensional models, only rotations by $\pi$ are allowed so the noise is implemented by randomly changing the sign of the velocity ($v\to-v$) with a certain rate $\lambda$. 
The dynamics of the velocity-flipping model are substantially different from that of the self-consistent reservoirs. Notwithstanding, it has been shown by Lukkarinen that both have the same steady-state \cite{Lukkarinen2012}. 
Finally, we mention that in quantum systems the noise is usually implemented using the Lindblad master equation. For a recent example involving quantum transport, see~\cite{Asadian2013}.

The rectification may be quantified as follows.
Let $J$ denote the heat flux over a given bath configuration and $J'$ the corresponding flux obtained by reversing the baths. 
We  define the rectification coefficient as
\begin{equation}\label{R}
R = \frac{|J|-|J'|}{|J| + |J'|}
\end{equation}
It is zero when there is no rectification and $\pm 1$ when one direction acts as a perfect heat insulator.

In this paper we will show that including an energy-conserving noise leads to a finite rectification effect in the thermodynamic limit. 
We will consider the problem of a classical  asymmetric harmonic chain with quartic ($\Phi^4$) pinning interaction and coupled to Langevin heat baths kept at different temperatures.
It will be shown that without the energy-conserving noise $R$ tends to zero in the thermodynamic limit. 
 But for any non-zero noise, $R$ tends to a  finite value. 

It is worth emphasizing that these numerical simulations are computationally very expensive, due mainly to three reasons. 
First, the rectification is much more sensitive to numerical fluctuations than the heat flux. 
Secondly, the energy-conserving noise requires the generation of a large number of random numbers.
Finally, in the presence of the energy-conserving noise the number of statistical averages required to obtain a good convergence increases substantially. 
Hence our choice of focusing on a specific model. 
Notwithstanding, we have also observed a similar behavior for other mechanical models, such as the Ferm-Pasta-Ulam and the Frenkel-Kontorova models. 
And in all cases the results are  unambiguous: The presence of the energy conserving noise leads to a finite rectification in the thermodynamic limit.

%----------------------------------------------------------
\section{Model}

We propose to study a chain of $L$ particles with unit masses
described by coordinates
$x_i$ and velocity $v_i$ associated to the following Hamiltonian 
\begin{equation}
H = \frac{1}{2} \sum_{i=1}^L  v_i^2
+ \frac{k}{2} \sum\limits_{i=0}^L (x_i-x_{i+1})^2
+ \frac{g_L}{4} \sum_{i=1}^{L/2} x_i^4  + \frac{g_R}{4} \sum_{i=\frac{L}{2}+1}^{L} x_i^4
\label{H}
\end{equation}
with fixed boundary conditions, $x_0=x_{L+1}=0$.
In addition to the nearest-neighbor harmonic interaction, the chain is also subject to a quartic pinning term which is different for the left and right halves of the chain. 
In this model the quartic pinning acts as both the asymmetry (when $g_L \neq g_R$) and the anharmonicity (to ensure a temperature dependence of the thermal conductivity).

The system is connected to two heat baths placed at the ends of the chain and modeled  by Langevin equations. 
The equations of motion are, therefore,
\begin{IEEEeqnarray}{rCl}
\label{lang1} m\frac{dv_1}{dt} &=& F_1  - \alpha v_1 + \sqrt{2\alpha T_A} \,\xi_1(t), \\[0.2cm]
\label{langi} m\frac{dv_i}{dt} &=& F_i, \qquad i = 2,\ldots,L-1, \\[0.2cm]
\label{langN} m\frac{dv_L}{dt} &=& F_L  - \alpha v_L + \sqrt{2\alpha T_B} \,\xi_L(t),
\end{IEEEeqnarray}
where $F_i = - {\partial{H}}/{\partial x_i}$ is the force acting on  particle $i$,
%given by
%\begin{equation}
%F_i = k(x_{i-1}+x_{i+1}-2x_i) - gx_i^3,
%\label{f1}
%\end{equation}
%for $1\leq i \leq L/2$, and
%\begin{equation}
%F_i = k(x_{i-1}+x_{i+1}-2x_i),
%\label{f2}
%\end{equation}
%for $L/2< i \leq L$; 
$\alpha$ is the damping constant and $\xi_i$ are independent standard Gaussian white noises. 
The temperatures of the two reservoirs, $A$ and $B$,
are denoted by $T_A$ and $T_B$ (and $k_B = 1$).

In addition to the baths, all particles are subject to an energy-conserving  noise that randomly flips the velocity of each particle
with a rate $\lambda$. 
The effects of this noise are best seen in terms of the master equation for the probability distribution $P(x,v,t)$, where $x=(x_1,\ldots,x_L)$ and $v=(v_1\ldots,v_L)$. 
It reads
\begin{equation}\label{FP}
\frac{\partial P}{\partial t} = {\cal L} P + \lambda \sum\limits_{i=1}^L \left[P (v^i) - P(v)\right]
\end{equation}
where  $v^i = (v_1,\ldots,-v_i,\ldots,v_L)$ and
\[
{\cal L}P = -\sum\limits_i \Bigg\{ \frac{1}{m} \frac{\partial}{\partial v_i}\left(F_i P - \alpha_i v_i P\right) - \frac{\partial (v_i P)}{\partial x_i}  + \frac{\alpha_i T_i }{m^2} \frac{\partial^2 P}{\partial v_i^2}  \Bigg\}
\]
is the usual Fokker-Planck operator containing the Langevin dynamics, with $\alpha_1 = \alpha_N = \alpha$ and $\alpha_i = 0$ otherwise. We also have $T_1 = T_A$ and $T_N = T_B$.
%Thus, as can be seen, the dynamics is governed by an equation which
%is a combination of a Fokker-Planck equation and a master equation. 

Since the noise conserves the total energy, the rate of change of $H$ should only be affected by the Langevin reservoirs. 
Indeed, using Eq.~(\ref{FP}) one arrives at 
\begin{equation}
\frac{d}{dt}\langle H \rangle = J_A - J_B
\end{equation}
where 
\begin{IEEEeqnarray}{rCl}
\label{JA} J_A &=& \alpha(T_A - \langle v_1^2 \rangle) \\[0.2cm]
\label{JB} J_B &=& -\alpha (T_B - \langle v_L^2 \rangle)
\end{IEEEeqnarray}
are the fluxes from the left and right baths respectively.
In the steady-state $J_A = J_B := J$.

It is worth mentioning that from a computational perspective, it is more efficient to compute the heat flux from the formula 
\begin{equation} 
J_i = k\langle v_i (x_{i} - x_{i+1}) \rangle 
\label{J}
\end{equation}
which corresponds to the heat from site $i$ to site $i+1$. The reason is that in Eqs.~(\ref{JA}) and (\ref{JB}),  quantities such as $T_A - \langle v_1^2 \rangle$ are extremely small  and therefore prone to numerical fluctuations. 

We also note that the existence of a heat current entails a continuous production of entropy. 
In the steady-state, the entropy production rate is given by \cite{tome2010}
\begin{equation}
\Pi = J\left(\frac1{T_B}-\frac1{T_A}\right).
\end{equation}
Thus, if both temperatures are equal, no entropy is produced, even in the presence of the energy-conserving noise.

%----------------------------------------------------------
\section{Numerical simulations}

The Langevin equations (\ref{lang1})-(\ref{langN})
were integrated numerically to obtain the heat flux in Eq.~(\ref{J}). For the inner particles we
used the velocity-Verlet algorithm \cite{Verlet1967} and for the boundary
particles we use the Stochastic Verlet algorithm developed in
\cite{Gronbech-Jensen2013}.
The action of the energy-conserving noise is as follows.
 Let $\Delta t$ be the time step
of the integration. 
For each particle, at each time step,  we flip the velocity of the particle
with probability $p = \lambda \Delta t$. This is a Poisson process which
generates a geometric distribution with discrete waiting time $p(1-p)^\ell$,
where $\ell$ is an integer. In the continuous limit this tends to an exponential
distribution $\lambda e^{-\lambda t}$. A more efficient implementation is the
following. For each particle, draw a random number $\tau_i$ from an exponential 
distribution with parameter $\lambda$. At each time step decrease each
$\tau_i$ by $\Delta t$. Reverse the velocity when $\tau_i$ becomes negative
and, afterwards, draw a new $\tau_i$. The results are of course the same,
but this requires the generation of less random numbers, while  allowing for
a larger $\Delta t$ to be used without loss of accuracy. 

In what follows we fix the spring constant  at $k=1$ and the damping term at $\alpha =0.1$. Moreover, to investigate the thermal rectification we consider 
$(T_A,T_B) = (1.8,0.2)$ and $(0.2,1.8)$. The corresponding fluxes will be referred to as $J$ and $J'$ respectively. 
Finally, as for the quartic pinning terms, we will consider two situations: $(g_L,g_R) = (0,5)$ and $(1,5)$. 
Note that in the first situation the left chain is  harmonic.

\begin{figure}[!h]
\centering
\includegraphics[width=0.45\textwidth]{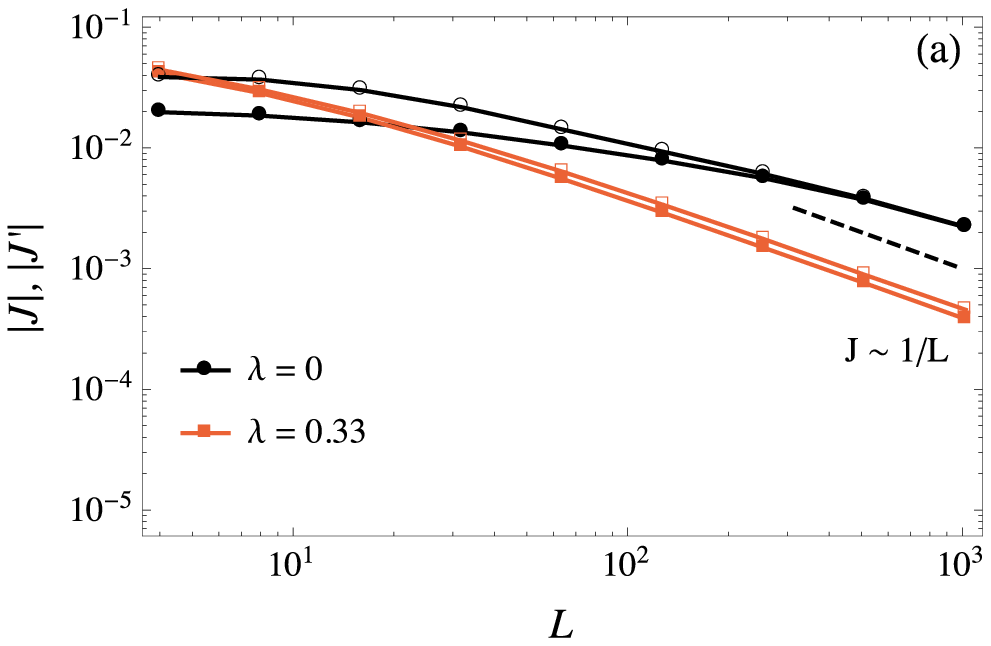}\\
\includegraphics[width=0.45\textwidth]{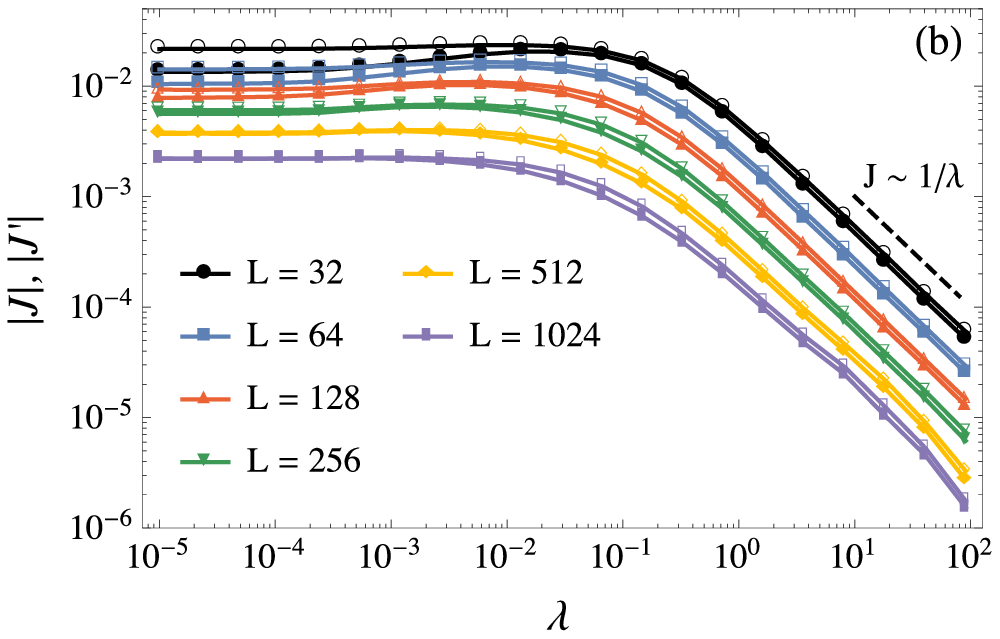}
\caption{\label{fig:J} 
(Color Online)
Heat fluxes $J$ (filled points) and $J'$ (open points) for $(g_L,g_R) = (0,5)$.
(a) As a function of $L$ for different values of $\lambda$.
(b) As a function of $\lambda$ for different values of $L$.
In both cases the dashed lines represent slope -1.
}
\end{figure}

We start with $(g_L,g_R) = (0,5)$.
Fig.~\ref{fig:J} shows both $J$ and $J'$ (filled and open points respectively)  as a function of $L$ and $\lambda$. 
In Fig.~\ref{fig:J}(a) we see that for sufficiently large $L$ the flux is diffusive, behaving as $J\sim 1/L$.
This is of course expected since both the quartic pinning \cite{Aoki2000a,*Aoki2006} and the energy-conserving  noise \cite{Dhar2011,Landi2014a,Landi2013a} are known to lead to Fourier's law. 
A similar behavior is also observed for $J$ as a function of $\lambda$, as  seen in Fig.~\ref{fig:J}(b). 
This is in agreement with what was found Ref.~\cite{Landi2014a,Landi2013a}.
Hence, we conclude that if both $\lambda$ and $L$ are sufficiently large, the flux behaves as 
\begin{equation}\label{Jscaling}
J =\frac{b}{\lambda L}
\end{equation}
with different coefficients $b$ and $b'$ for $J$ and $J'$ respectively.

The existence of rectification and the role played by the energy conserving noise may be visualized quite clearly in Fig.~\ref{fig:J}(a). 
When $\lambda=0$ the fluxes $J$ and $J'$ are first distinct when $L$ is small, but gradually approach one another as $L$ becomes large.
This is the behavior found in most models of thermal rectification. 
And it shows clearly that the rectification vanishes in the thermodynamic limit.
Conversely, when $\lambda$ is large, a finite separation between $J$ and $J'$ remains for any size considered. 
Hence, for any non-zero $\lambda$ the rectification remains \emph{finite} in the thermodynamic limit.

\begin{figure}
\centering
\includegraphics[width=0.45\textwidth]{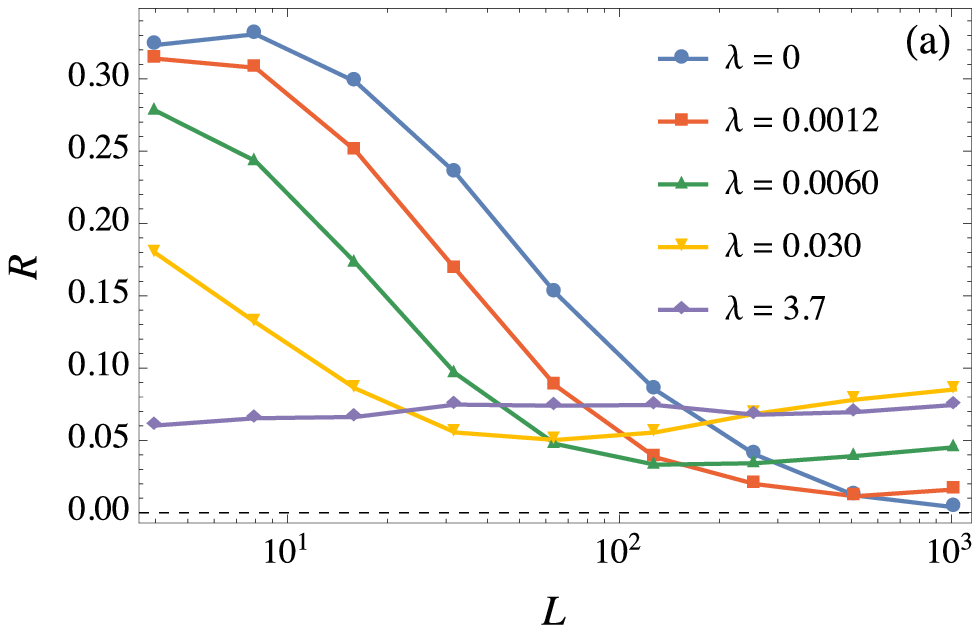}\\
\includegraphics[width=0.45\textwidth]{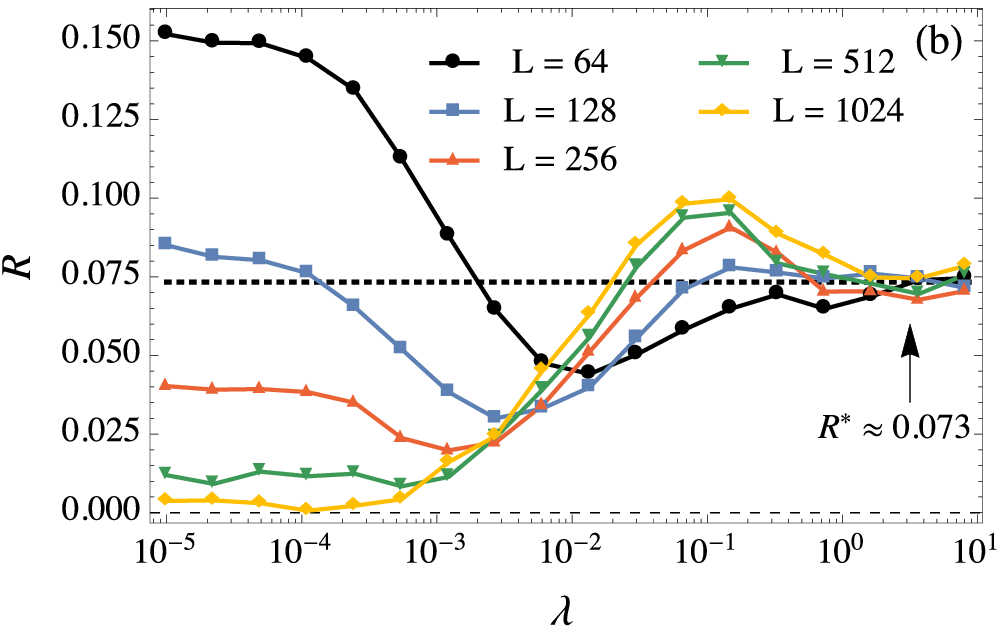}
\caption{\label{fig:R} (Color Online)
Rectification coefficient $R$ computed from Eq.~(\ref{R}) and the data of Fig.~\ref{fig:J}.
(a) As a function of $L$ for different values of $\lambda$.
(b) As a function of $\lambda$ for different values of $L$. 
The dotted line in (b) corresponds to the value $R^* = 0.073$ which the curves for all sizes tend to when $\lambda$ is sufficiently large.
}
\end{figure}

A more quantitative analysis may be done by computing the rectification coefficient in Eq.~(\ref{R}).
The results are shown in Fig.~\ref{fig:R}. 
It is important to mention that accurately computing $R$  is an extremely challenging task. 
First, it involves taking the difference between two fluctuating quantities. 
Secondly, since $J\propto 1/L$, the fluxes are already quite small when $L$ is large.
Finally, the statistical fluctuations increase substantially with $\lambda$.

Let us first analyze   Fig.~\ref{fig:R}(a), where $R$ is plotted as a function of $L$ for several values of $\lambda$.  
When $\lambda = 0$ the rectification is initially large, but then goes down with $L$ and tends to zero in the thermodynamic limit. 
As $\lambda$ increases, however, one observes that for large $L$ the rectification settles at a non-zero value. 
If $\lambda$ is sufficiently large  the rectification  even becomes roughly independent of size.

Similar conclusions may be drawn from Fig.~\ref{fig:R}(b), where we plot $R$ as a function of $\lambda$ for several sizes. 
If we look at the low $\lambda$ region we see that $R$ diminishes quickly from $R\sim 0.15$ to zero as $L$ increases. 
But for large $\lambda$ all curves are found to approach a common value of roughly $R^* \simeq 0.073$. 

\begin{figure}[!h]
\centering
\includegraphics[width=0.45\textwidth]{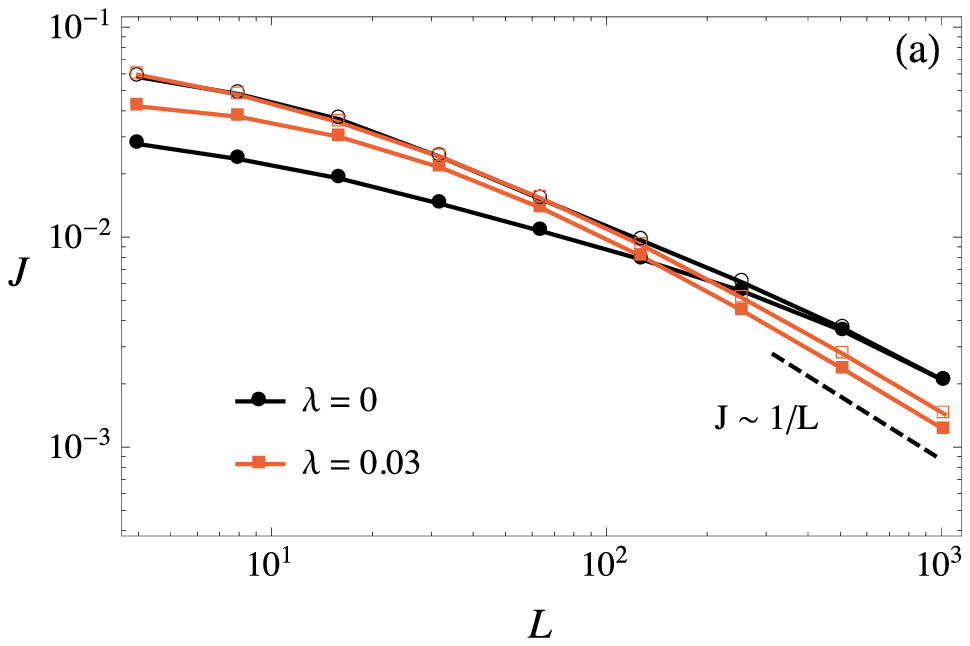}\\
\includegraphics[width=0.45\textwidth]{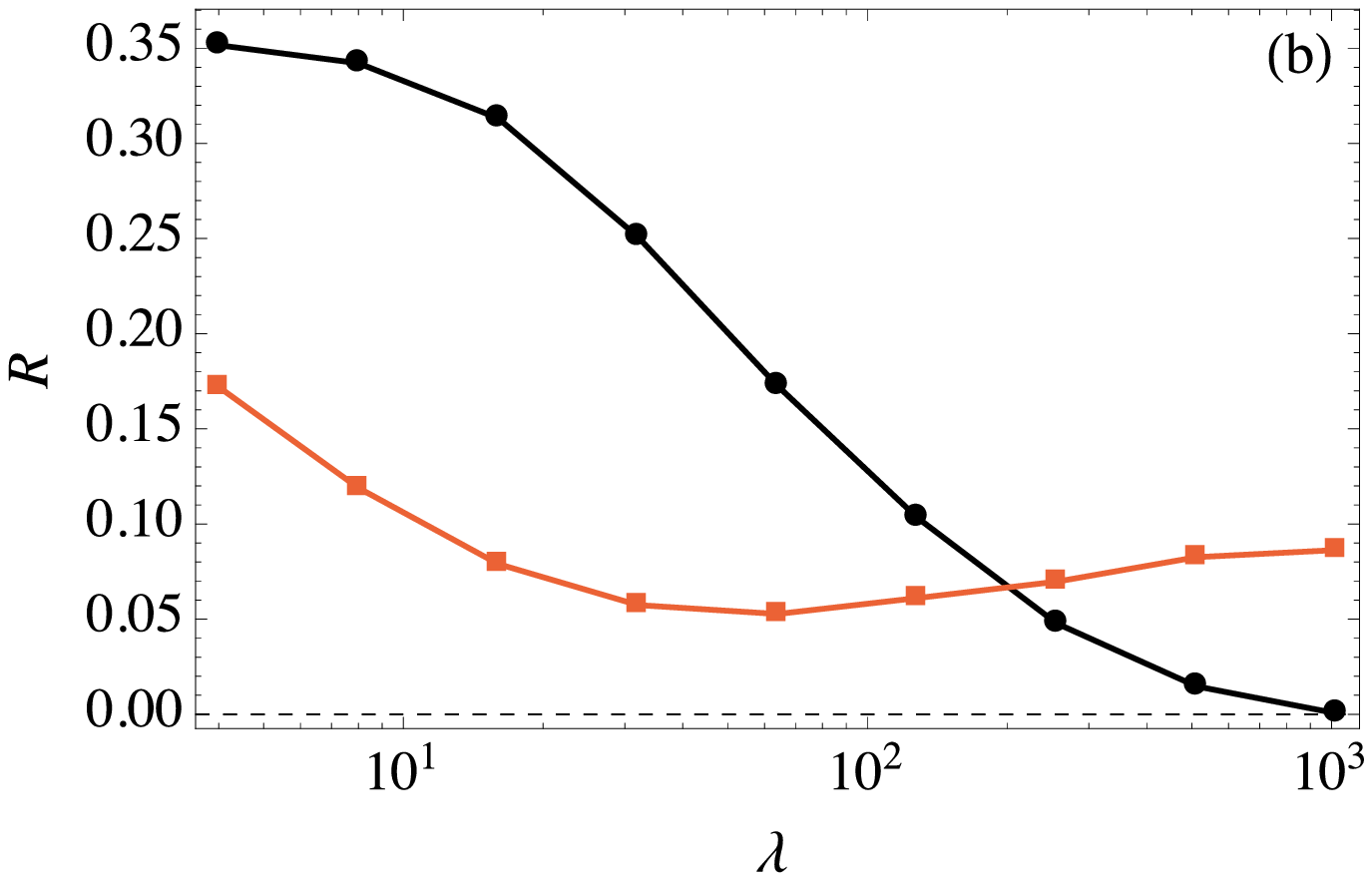}
\caption{\label{fig:p2} 
(Color Online)
(a) Heat fluxes and (b) thermal rectification as a function of $L$, for $(g_L,g_R) = (1,5)$ and two  values of $\lambda$. 
}
\end{figure}

In the situation considered so far, the left part of the chain was actually harmonic, since there is no pinning term ($g_L = 0$). 
To show that this does not influence the main effect of the energy-conserving noise, we now consider a condition in which $g_L \neq 0$. 
In Fig.~\ref{fig:p2} we plot the fluxes and the corresponding rectification coefficient for $(g_L,g_R) = (1,5)$.
As can be seen, the conclusions are all identical to the previous case: when $\lambda = 0$ (black circles) the rectification clearly tends to zero as $L$ increases. 
Conversely, when $\lambda = 0.03$ (red squares), the rectification remains finite for any size.  
%This is relevant because in the previous case the left part of the chain was harmonic $(g_L = 0)$. 
%Thus, this agreement shows that the inclusion of the energy-conserving noise leads to similar results irrespective of whether the chain is harmonic or not. 

%----------------------------------------------------------
\section{Discussion and Conclusions}

The rectification effect in the studied models exist for both with ($\lambda \neq 0$) and without ($\lambda = 0$) the energy-conserving noise.
But their dependence on the size $L$ of the system is entirely different in either case. 
Note that even when $\lambda = 0$ the model presents a diffusive heat flux so to leading order we may write 
\[
|J| = \frac{A}{L},\qquad |J'| = \frac{A'}{L},
\]
which, from Eq.~(\ref{R}), implies a rectification coefficient 
\[
R = \frac{A-A'}{A+A'}.
\]
Our results show that when $\lambda = 0$, $A'=A$ and hence the rectification is zero. 
The existence of a non-zero energy-conserving noise therefore ensures a finite rectification. 

When $\lambda=0$ we may have, for instance, a situation where
\[
|J| = \frac{A}{L} + \frac{B}{L^2},\qquad |J'| = \frac{A}{L} + \frac{B'}{L^2},
\]
which, for large $L$, would lead to a rectification coefficient 
\[
R = \frac{B-B'}{2A L}.
\]
Since $R\propto1/L$, the effect becomes negligible in the thermodynamic limit.

Thermal rectification has been measured experimentally in macroscopic systems \cite{Starr1936,Brattain1951,Chang2006,Scheibner2008,Hu2009,Kobayashi2009,Yang2009,Nefzaoui2014},  which is
understood as a very large system from the microscopic viewpoint.
Thus, if we wish to compare the results of a microscopic
model with experiment, we uderstand that the thermodynamic limit
should be taken. 
However, in most theoretical models, the magnitude of the effect vanishes in the thermodynamic limit.

The purpose of this paper was to show that  including an energy-conserving noise leads to a finite rectification in the thermodynamic limit.
To illustrate this  we studied  the problem of two classical anharmonic chains with different pinning coefficients. 
But we have also studied other classical models, such as the Frenkel-Kontorova  and Fermi-Pasta-Ulam  (results not shown).
In all cases, the results are unambiguous: a non-zero noise leads to a finite rectification in the thermodynamic limit. 
These results therefore lead to the conjecture that the existence of a finite rectification in the presence of the energy conserving noise may be a universal behavior, independent of the underlaying mechanical model. 
Lastly, we mention that the physical explanation of this effect, although not trivial, is certainly related to the highly chaotic behavior of the energy-conserving noise, 
which makes the system ergodic \cite{Landi2013a}.

%----------------------------------------------------------
\section{Acknowledgement}

The authors acknowledge the S\~ao Paulo Research Foundation (FAPESP) and CNPq for the financial support.

\bibliography{/Users/gtlandi/Documents/library}

\begin{thebibliography}{48}%
\makeatletter
\providecommand \@ifxundefined [1]{%
 \@ifx{#1\undefined}
}%
\providecommand \@ifnum [1]{%
 \ifnum #1\expandafter \@firstoftwo
 \else \expandafter \@secondoftwo
 \fi
}%
\providecommand \@ifx [1]{%
 \ifx #1\expandafter \@firstoftwo
 \else \expandafter \@secondoftwo
 \fi
}%
\providecommand \natexlab [1]{#1}%
\providecommand \enquote  [1]{``#1''}%
\providecommand \bibnamefont  [1]{#1}%
\providecommand \bibfnamefont [1]{#1}%
\providecommand \citenamefont [1]{#1}%
\providecommand \href@noop [0]{\@secondoftwo}%
\providecommand \href [0]{\begingroup \@sanitize@url \@href}%
\providecommand \@href[1]{\@@startlink{#1}\@@href}%
\providecommand \@@href[1]{\endgroup#1\@@endlink}%
\providecommand \@sanitize@url [0]{\catcode `\\12\catcode `\$12\catcode
  `\&12\catcode `\#12\catcode `\^12\catcode `\_12\catcode `\%12\relax}%
\providecommand \@@startlink[1]{}%
\providecommand \@@endlink[0]{}%
\providecommand \url  [0]{\begingroup\@sanitize@url \@url }%
\providecommand \@url [1]{\endgroup\@href {#1}{\urlprefix }}%
\providecommand \urlprefix  [0]{URL }%
\providecommand \Eprint [0]{\href }%
\providecommand \doibase [0]{http://dx.doi.org/}%
\providecommand \selectlanguage [0]{\@gobble}%
\providecommand \bibinfo  [0]{\@secondoftwo}%
\providecommand \bibfield  [0]{\@secondoftwo}%
\providecommand \translation [1]{[#1]}%
\providecommand \BibitemOpen [0]{}%
\providecommand \bibitemStop [0]{}%
\providecommand \bibitemNoStop [0]{.\EOS\space}%
\providecommand \EOS [0]{\spacefactor3000\relax}%
\providecommand \BibitemShut  [1]{\csname bibitem#1\endcsname}%
\let\auto@bib@innerbib\@empty
%</preamble>
\bibitem [{\citenamefont {Hu}\ \emph {et~al.}(2006)\citenamefont {Hu},
  \citenamefont {Yang},\ and\ \citenamefont {Zhang}}]{Hu2006}%
  \BibitemOpen
  \bibfield  {author} {\bibinfo {author} {\bibfnamefont {B.}~\bibnamefont
  {Hu}}, \bibinfo {author} {\bibfnamefont {L.}~\bibnamefont {Yang}}, \ and\
  \bibinfo {author} {\bibfnamefont {Y.}~\bibnamefont {Zhang}},\ }\href
  {\doibase 10.1103/PhysRevLett.97.124302} {\bibfield  {journal} {\bibinfo
  {journal} {Physical Review Letters}\ }\textbf {\bibinfo {volume} {97}},\
  \bibinfo {pages} {124302} (\bibinfo {year} {2006})}\BibitemShut {NoStop}%
\bibitem [{\citenamefont {Terraneo}\ \emph {et~al.}(2002)\citenamefont
  {Terraneo}, \citenamefont {Peyrard},\ and\ \citenamefont
  {Casati}}]{Terraneo2002}%
  \BibitemOpen
  \bibfield  {author} {\bibinfo {author} {\bibfnamefont {M.}~\bibnamefont
  {Terraneo}}, \bibinfo {author} {\bibfnamefont {M.}~\bibnamefont {Peyrard}}, \
  and\ \bibinfo {author} {\bibfnamefont {G.}~\bibnamefont {Casati}},\ }\href
  {\doibase 10.1103/PhysRevLett.88.094302} {\bibfield  {journal} {\bibinfo
  {journal} {Physical Review Letters}\ }\textbf {\bibinfo {volume} {88}},\
  \bibinfo {pages} {094302} (\bibinfo {year} {2002})}\BibitemShut {NoStop}%
\bibitem [{\citenamefont {Li}\ \emph {et~al.}(2004)\citenamefont {Li},
  \citenamefont {Wang},\ and\ \citenamefont {Casati}}]{Li2004a}%
  \BibitemOpen
  \bibfield  {author} {\bibinfo {author} {\bibfnamefont {B.}~\bibnamefont
  {Li}}, \bibinfo {author} {\bibfnamefont {L.}~\bibnamefont {Wang}}, \ and\
  \bibinfo {author} {\bibfnamefont {G.}~\bibnamefont {Casati}},\ }\href
  {\doibase 10.1103/PhysRevLett.93.184301} {\bibfield  {journal} {\bibinfo
  {journal} {Physical Review Letters}\ }\textbf {\bibinfo {volume} {93}},\
  \bibinfo {pages} {184301} (\bibinfo {year} {2004})}\BibitemShut {NoStop}%
\bibitem [{\citenamefont {Pereira}\ and\ \citenamefont
  {Falcao}(2004)}]{Pereira2004}%
  \BibitemOpen
  \bibfield  {author} {\bibinfo {author} {\bibfnamefont {E.}~\bibnamefont
  {Pereira}}\ and\ \bibinfo {author} {\bibfnamefont {R.}~\bibnamefont
  {Falcao}},\ }\href {\doibase 10.1103/PhysRevE.70.046105} {\bibfield
  {journal} {\bibinfo  {journal} {Physical Review E}\ }\textbf {\bibinfo
  {volume} {70}},\ \bibinfo {pages} {046105} (\bibinfo {year}
  {2004})}\BibitemShut {NoStop}%
\bibitem [{\citenamefont {Li}\ \emph {et~al.}(2005)\citenamefont {Li},
  \citenamefont {Lan},\ and\ \citenamefont {Wang}}]{Li2005}%
  \BibitemOpen
  \bibfield  {author} {\bibinfo {author} {\bibfnamefont {B.}~\bibnamefont
  {Li}}, \bibinfo {author} {\bibfnamefont {J.}~\bibnamefont {Lan}}, \ and\
  \bibinfo {author} {\bibfnamefont {L.}~\bibnamefont {Wang}},\ }\href {\doibase
  10.1103/PhysRevLett.95.104302} {\bibfield  {journal} {\bibinfo  {journal}
  {Physical Review Letters}\ }\textbf {\bibinfo {volume} {95}},\ \bibinfo
  {pages} {104302} (\bibinfo {year} {2005})}\BibitemShut {NoStop}%
\bibitem [{\citenamefont {Wang}\ and\ \citenamefont {Li}(2007)}]{Wang2007}%
  \BibitemOpen
  \bibfield  {author} {\bibinfo {author} {\bibfnamefont {L.}~\bibnamefont
  {Wang}}\ and\ \bibinfo {author} {\bibfnamefont {B.}~\bibnamefont {Li}},\
  }\href {\doibase 10.1103/PhysRevLett.99.177208} {\bibfield  {journal}
  {\bibinfo  {journal} {Physical Review Letters}\ }\textbf {\bibinfo {volume}
  {99}},\ \bibinfo {pages} {177208} (\bibinfo {year} {2007})}\BibitemShut
  {NoStop}%
\bibitem [{\citenamefont {Lo}\ \emph {et~al.}(2008)\citenamefont {Lo},
  \citenamefont {Wang},\ and\ \citenamefont {Li}}]{Lo2008}%
  \BibitemOpen
  \bibfield  {author} {\bibinfo {author} {\bibfnamefont {W.~C.}\ \bibnamefont
  {Lo}}, \bibinfo {author} {\bibfnamefont {L.}~\bibnamefont {Wang}}, \ and\
  \bibinfo {author} {\bibfnamefont {B.}~\bibnamefont {Li}},\ }\href@noop {}
  {\bibfield  {journal} {\bibinfo  {journal} {Journal of the Physical Society
  of Japan}\ }\textbf {\bibinfo {volume} {77}},\ \bibinfo {pages} {054402}
  (\bibinfo {year} {2008})}\BibitemShut {NoStop}%
\bibitem [{\citenamefont {Pereira}\ and\ \citenamefont
  {Lemos}(2008)}]{Pereira2008}%
  \BibitemOpen
  \bibfield  {author} {\bibinfo {author} {\bibfnamefont {E.}~\bibnamefont
  {Pereira}}\ and\ \bibinfo {author} {\bibfnamefont {H.~C.~F.}\ \bibnamefont
  {Lemos}},\ }\href {\doibase 10.1103/PhysRevE.78.031108} {\bibfield  {journal}
  {\bibinfo  {journal} {Physical Review E}\ }\textbf {\bibinfo {volume} {78}},\
  \bibinfo {pages} {031108} (\bibinfo {year} {2008})}\BibitemShut {NoStop}%
\bibitem [{\citenamefont {Scheibner}\ \emph {et~al.}(2008)\citenamefont
  {Scheibner}, \citenamefont {K\"{o}nig}, \citenamefont {Reuter}, \citenamefont
  {Wieck}, \citenamefont {Gould}, \citenamefont {Buhmann},\ and\ \citenamefont
  {Molenkamp}}]{Scheibner2008}%
  \BibitemOpen
  \bibfield  {author} {\bibinfo {author} {\bibfnamefont {R.}~\bibnamefont
  {Scheibner}}, \bibinfo {author} {\bibfnamefont {M.}~\bibnamefont
  {K\"{o}nig}}, \bibinfo {author} {\bibfnamefont {D.}~\bibnamefont {Reuter}},
  \bibinfo {author} {\bibfnamefont {a.~D.}\ \bibnamefont {Wieck}}, \bibinfo
  {author} {\bibfnamefont {C.}~\bibnamefont {Gould}}, \bibinfo {author}
  {\bibfnamefont {H.}~\bibnamefont {Buhmann}}, \ and\ \bibinfo {author}
  {\bibfnamefont {L.~W.}\ \bibnamefont {Molenkamp}},\ }\href {\doibase
  10.1088/1367-2630/10/8/083016} {\bibfield  {journal} {\bibinfo  {journal}
  {New Journal of Physics}\ }\textbf {\bibinfo {volume} {10}},\ \bibinfo
  {pages} {083016} (\bibinfo {year} {2008})}\BibitemShut {NoStop}%
\bibitem [{\citenamefont {Hu}\ \emph {et~al.}(2009)\citenamefont {Hu},
  \citenamefont {Ruan},\ and\ \citenamefont {Chen}}]{Hu2009}%
  \BibitemOpen
  \bibfield  {author} {\bibinfo {author} {\bibfnamefont {J.}~\bibnamefont
  {Hu}}, \bibinfo {author} {\bibfnamefont {X.}~\bibnamefont {Ruan}}, \ and\
  \bibinfo {author} {\bibfnamefont {Y.~P.}\ \bibnamefont {Chen}},\ }\href
  {http://pubs.acs.org/doi/abs/10.1021/nl901231s} {\bibfield  {journal}
  {\bibinfo  {journal} {Nano Letters}\ }\textbf {\bibinfo {volume} {9}},\
  \bibinfo {pages} {2730} (\bibinfo {year} {2009})}\BibitemShut {NoStop}%
\bibitem [{\citenamefont {Kobayashi}\ \emph {et~al.}(2009)\citenamefont
  {Kobayashi}, \citenamefont {Teraoka},\ and\ \citenamefont
  {Terasaki}}]{Kobayashi2009}%
  \BibitemOpen
  \bibfield  {author} {\bibinfo {author} {\bibfnamefont {W.}~\bibnamefont
  {Kobayashi}}, \bibinfo {author} {\bibfnamefont {Y.}~\bibnamefont {Teraoka}},
  \ and\ \bibinfo {author} {\bibfnamefont {I.}~\bibnamefont {Terasaki}},\
  }\href {\doibase 10.1063/1.3253712} {\bibfield  {journal} {\bibinfo
  {journal} {Applied Physics Letters}\ }\textbf {\bibinfo {volume} {95}},\
  \bibinfo {pages} {171905} (\bibinfo {year} {2009})}\BibitemShut {NoStop}%
\bibitem [{\citenamefont {Yang}\ \emph {et~al.}(2009)\citenamefont {Yang},
  \citenamefont {Zhang},\ and\ \citenamefont {Li}}]{Yang2009}%
  \BibitemOpen
  \bibfield  {author} {\bibinfo {author} {\bibfnamefont {N.}~\bibnamefont
  {Yang}}, \bibinfo {author} {\bibfnamefont {G.}~\bibnamefont {Zhang}}, \ and\
  \bibinfo {author} {\bibfnamefont {B.}~\bibnamefont {Li}},\ }\href {\doibase
  10.1063/1.3183587} {\bibfield  {journal} {\bibinfo  {journal} {Applied
  Physics Letters}\ }\textbf {\bibinfo {volume} {95}},\ \bibinfo {pages}
  {033107} (\bibinfo {year} {2009})}\BibitemShut {NoStop}%
\bibitem [{\citenamefont {Zhang}\ \emph {et~al.}(2009)\citenamefont {Zhang},
  \citenamefont {Yan}, \citenamefont {Wu}, \citenamefont {Wang},\ and\
  \citenamefont {Li}}]{Zhang2009}%
  \BibitemOpen
  \bibfield  {author} {\bibinfo {author} {\bibfnamefont {L.}~\bibnamefont
  {Zhang}}, \bibinfo {author} {\bibfnamefont {Y.}~\bibnamefont {Yan}}, \bibinfo
  {author} {\bibfnamefont {C.-Q.}\ \bibnamefont {Wu}}, \bibinfo {author}
  {\bibfnamefont {J.-S.}\ \bibnamefont {Wang}}, \ and\ \bibinfo {author}
  {\bibfnamefont {B.}~\bibnamefont {Li}},\ }\href {\doibase
  10.1103/PhysRevB.80.172301} {\bibfield  {journal} {\bibinfo  {journal}
  {Physical Review B}\ }\textbf {\bibinfo {volume} {80}},\ \bibinfo {pages}
  {172301} (\bibinfo {year} {2009})}\BibitemShut {NoStop}%
\bibitem [{\citenamefont {Pereira}(2010{\natexlab{a}})}]{Pereira2010b}%
  \BibitemOpen
  \bibfield  {author} {\bibinfo {author} {\bibfnamefont {E.}~\bibnamefont
  {Pereira}},\ }\href {\doibase 10.1016/j.physleta.2010.02.071} {\bibfield
  {journal} {\bibinfo  {journal} {Physics Letters A}\ }\textbf {\bibinfo
  {volume} {374}},\ \bibinfo {pages} {1933} (\bibinfo {year}
  {2010}{\natexlab{a}})}\BibitemShut {NoStop}%
\bibitem [{\citenamefont {Pereira}(2010{\natexlab{b}})}]{Pereira2010}%
  \BibitemOpen
  \bibfield  {author} {\bibinfo {author} {\bibfnamefont {E.}~\bibnamefont
  {Pereira}},\ }\href {\doibase 10.1103/PhysRevE.82.040101} {\bibfield
  {journal} {\bibinfo  {journal} {Physical Review E}\ }\textbf {\bibinfo
  {volume} {82}},\ \bibinfo {pages} {040101} (\bibinfo {year}
  {2010}{\natexlab{b}})}\BibitemShut {NoStop}%
\bibitem [{\citenamefont {Pereira}(2011)}]{Pereira2011}%
  \BibitemOpen
  \bibfield  {author} {\bibinfo {author} {\bibfnamefont {E.}~\bibnamefont
  {Pereira}},\ }\href {\doibase 10.1103/PhysRevE.83.031106} {\bibfield
  {journal} {\bibinfo  {journal} {Physical Review E}\ }\textbf {\bibinfo
  {volume} {83}},\ \bibinfo {pages} {031106} (\bibinfo {year}
  {2011})}\BibitemShut {NoStop}%
\bibitem [{\citenamefont {Pereira}\ \emph {et~al.}(2011)\citenamefont
  {Pereira}, \citenamefont {Lemos},\ and\ \citenamefont
  {\'{A}vila}}]{Pereira2011b}%
  \BibitemOpen
  \bibfield  {author} {\bibinfo {author} {\bibfnamefont {E.}~\bibnamefont
  {Pereira}}, \bibinfo {author} {\bibfnamefont {H.~C.~F.}\ \bibnamefont
  {Lemos}}, \ and\ \bibinfo {author} {\bibfnamefont {R.~R.}\ \bibnamefont
  {\'{A}vila}},\ }\href {\doibase 10.1103/PhysRevE.84.061135} {\bibfield
  {journal} {\bibinfo  {journal} {Physical Review E}\ }\textbf {\bibinfo
  {volume} {84}},\ \bibinfo {pages} {061135} (\bibinfo {year}
  {2011})}\BibitemShut {NoStop}%
\bibitem [{\citenamefont {Shah}\ and\ \citenamefont {Gajjar}(2012)}]{Shah2012}%
  \BibitemOpen
  \bibfield  {author} {\bibinfo {author} {\bibfnamefont {T.~N.}\ \bibnamefont
  {Shah}}\ and\ \bibinfo {author} {\bibfnamefont {P.}~\bibnamefont {Gajjar}},\
  }\href {\doibase 10.1016/j.physleta.2011.11.052} {\bibfield  {journal}
  {\bibinfo  {journal} {Physics Letters A}\ }\textbf {\bibinfo {volume}
  {376}},\ \bibinfo {pages} {438} (\bibinfo {year} {2012})}\BibitemShut
  {NoStop}%
\bibitem [{\citenamefont {Wang}\ \emph {et~al.}(2012)\citenamefont {Wang},
  \citenamefont {Pereira},\ and\ \citenamefont {Casati}}]{Wang2012a}%
  \BibitemOpen
  \bibfield  {author} {\bibinfo {author} {\bibfnamefont {J.}~\bibnamefont
  {Wang}}, \bibinfo {author} {\bibfnamefont {E.}~\bibnamefont {Pereira}}, \
  and\ \bibinfo {author} {\bibfnamefont {G.}~\bibnamefont {Casati}},\ }\href
  {\doibase 10.1103/PhysRevE.86.010101} {\bibfield  {journal} {\bibinfo
  {journal} {Physical Review E}\ }\textbf {\bibinfo {volume} {86}},\ \bibinfo
  {pages} {010101(R)} (\bibinfo {year} {2012})}\BibitemShut {NoStop}%
\bibitem [{\citenamefont {Pereira}\ and\ \citenamefont
  {\'{A}vila}(2013)}]{Pereira2013a}%
  \BibitemOpen
  \bibfield  {author} {\bibinfo {author} {\bibfnamefont {E.}~\bibnamefont
  {Pereira}}\ and\ \bibinfo {author} {\bibfnamefont {R.~R.}\ \bibnamefont
  {\'{A}vila}},\ }\href {\doibase 10.1103/PhysRevE.88.032139} {\bibfield
  {journal} {\bibinfo  {journal} {Physical Review E}\ }\textbf {\bibinfo
  {volume} {88}},\ \bibinfo {pages} {032139} (\bibinfo {year}
  {2013})}\BibitemShut {NoStop}%
\bibitem [{\citenamefont {Thingna}\ and\ \citenamefont
  {Wang}(2013)}]{Thingna2013}%
  \BibitemOpen
  \bibfield  {author} {\bibinfo {author} {\bibfnamefont {J.}~\bibnamefont
  {Thingna}}\ and\ \bibinfo {author} {\bibfnamefont {J.-S.}\ \bibnamefont
  {Wang}},\ }\href {\doibase 10.1209/0295-5075/104/37006} {\bibfield  {journal}
  {\bibinfo  {journal} {EPL (Europhysics Letters)}\ }\textbf {\bibinfo {volume}
  {104}},\ \bibinfo {pages} {37006} (\bibinfo {year} {2013})}\BibitemShut
  {NoStop}%
\bibitem [{\citenamefont {\'{A}vila}\ and\ \citenamefont
  {Pereira}(2013)}]{Avila2013}%
  \BibitemOpen
  \bibfield  {author} {\bibinfo {author} {\bibfnamefont {R.~R.}\ \bibnamefont
  {\'{A}vila}}\ and\ \bibinfo {author} {\bibfnamefont {E.}~\bibnamefont
  {Pereira}},\ }\href {\doibase 10.1088/1751-8113/46/5/055002} {\bibfield
  {journal} {\bibinfo  {journal} {Journal of Physics A: Mathematical and
  Theoretical}\ }\textbf {\bibinfo {volume} {46}},\ \bibinfo {pages} {055002}
  (\bibinfo {year} {2013})}\BibitemShut {NoStop}%
\bibitem [{\citenamefont {Romero-Bastida}\ and\ \citenamefont
  {Arizmendi-Carvajal}(2013)}]{Romero-Bastida2013}%
  \BibitemOpen
  \bibfield  {author} {\bibinfo {author} {\bibfnamefont {M.}~\bibnamefont
  {Romero-Bastida}}\ and\ \bibinfo {author} {\bibfnamefont {J.~M.}\
  \bibnamefont {Arizmendi-Carvajal}},\ }\href {\doibase
  10.1088/1751-8113/46/11/115006} {\bibfield  {journal} {\bibinfo  {journal}
  {Journal of Physics A: Mathematical and Theoretical}\ }\textbf {\bibinfo
  {volume} {46}},\ \bibinfo {pages} {115006} (\bibinfo {year}
  {2013})}\BibitemShut {NoStop}%
\bibitem [{\citenamefont {Werlang}\ \emph {et~al.}(2014)\citenamefont
  {Werlang}, \citenamefont {Marchiori}, \citenamefont {Cornelio},\ and\
  \citenamefont {Valente}}]{Werlang2014}%
  \BibitemOpen
  \bibfield  {author} {\bibinfo {author} {\bibfnamefont {T.}~\bibnamefont
  {Werlang}}, \bibinfo {author} {\bibfnamefont {M.~A.}\ \bibnamefont
  {Marchiori}}, \bibinfo {author} {\bibfnamefont {M.~F.}\ \bibnamefont
  {Cornelio}}, \ and\ \bibinfo {author} {\bibfnamefont {D.}~\bibnamefont
  {Valente}},\ }\href@noop {} {\bibfield  {journal} {\bibinfo  {journal}
  {arXiv}\ } (\bibinfo {year} {2014})},\ \Eprint
  {http://arxiv.org/abs/1404.0348v2} {arXiv:1404.0348v2} \BibitemShut {NoStop}%
\bibitem [{\citenamefont {Kubytskyi}\ \emph {et~al.}(2014)\citenamefont
  {Kubytskyi}, \citenamefont {Biehs},\ and\ \citenamefont
  {Ben-Abdallah}}]{Kubytskyi2014}%
  \BibitemOpen
  \bibfield  {author} {\bibinfo {author} {\bibfnamefont {V.}~\bibnamefont
  {Kubytskyi}}, \bibinfo {author} {\bibfnamefont {S.-A.}\ \bibnamefont
  {Biehs}}, \ and\ \bibinfo {author} {\bibfnamefont {P.}~\bibnamefont
  {Ben-Abdallah}},\ }\href {\doibase 10.1103/PhysRevLett.113.074301} {\bibfield
   {journal} {\bibinfo  {journal} {Physical Review Letters}\ }\textbf {\bibinfo
  {volume} {113}},\ \bibinfo {pages} {074301} (\bibinfo {year}
  {2014})}\BibitemShut {NoStop}%
\bibitem [{\citenamefont {Nefzaoui}\ \emph {et~al.}(2014)\citenamefont
  {Nefzaoui}, \citenamefont {Joulain}, \citenamefont {Drevillon},\ and\
  \citenamefont {Ezzahri}}]{Nefzaoui2014}%
  \BibitemOpen
  \bibfield  {author} {\bibinfo {author} {\bibfnamefont {E.}~\bibnamefont
  {Nefzaoui}}, \bibinfo {author} {\bibfnamefont {K.}~\bibnamefont {Joulain}},
  \bibinfo {author} {\bibfnamefont {J.}~\bibnamefont {Drevillon}}, \ and\
  \bibinfo {author} {\bibfnamefont {Y.}~\bibnamefont {Ezzahri}},\ }\href
  {\doibase 10.1063/1.4868251} {\bibfield  {journal} {\bibinfo  {journal}
  {Applied Physics Letters}\ }\textbf {\bibinfo {volume} {104}},\ \bibinfo
  {pages} {103905} (\bibinfo {year} {2014})}\BibitemShut {NoStop}%
\bibitem [{\citenamefont {Landi}\ \emph {et~al.}(2014)\citenamefont {Landi},
  \citenamefont {Novais}, \citenamefont {de~Oliveira},\ and\ \citenamefont
  {Karevski}}]{Landi2014b}%
  \BibitemOpen
  \bibfield  {author} {\bibinfo {author} {\bibfnamefont {G.~T.}\ \bibnamefont
  {Landi}}, \bibinfo {author} {\bibfnamefont {E.}~\bibnamefont {Novais}},
  \bibinfo {author} {\bibfnamefont {M.~J.}\ \bibnamefont {de~Oliveira}}, \ and\
  \bibinfo {author} {\bibfnamefont {D.}~\bibnamefont {Karevski}},\ }\href
  {\doibase 10.1103/PhysRevE.90.042142} {\bibfield  {journal} {\bibinfo
  {journal} {Physical Review E}\ }\textbf {\bibinfo {volume} {90}},\ \bibinfo
  {pages} {042142} (\bibinfo {year} {2014})}\BibitemShut {NoStop}%
\bibitem [{\citenamefont {Landi}\ and\ \citenamefont
  {Karevski}(2015)}]{Landi2015a}%
  \BibitemOpen
  \bibfield  {author} {\bibinfo {author} {\bibfnamefont {G.~T.}\ \bibnamefont
  {Landi}}\ and\ \bibinfo {author} {\bibfnamefont {D.}~\bibnamefont
  {Karevski}},\ }\href {\doibase 10.1103/PhysRevB.91.174422} {\bibfield
  {journal} {\bibinfo  {journal} {Physical Review B}\ }\textbf {\bibinfo
  {volume} {91}},\ \bibinfo {pages} {174422} (\bibinfo {year}
  {2015})}\BibitemShut {NoStop}%
\bibitem [{\citenamefont {Roberts}\ and\ \citenamefont
  {Walker}(2011)}]{Roberts2011}%
  \BibitemOpen
  \bibfield  {author} {\bibinfo {author} {\bibfnamefont {N.~A.}\ \bibnamefont
  {Roberts}}\ and\ \bibinfo {author} {\bibfnamefont {D.~G.}\ \bibnamefont
  {Walker}},\ }\href {\doibase 10.1016/j.ijthermalsci.2010.12.004} {\bibfield
  {journal} {\bibinfo  {journal} {International Journal of Thermal Sciences}\
  }\textbf {\bibinfo {volume} {50}},\ \bibinfo {pages} {648} (\bibinfo {year}
  {2011})}\BibitemShut {NoStop}%
\bibitem [{\citenamefont {Li}\ \emph {et~al.}(2012)\citenamefont {Li},
  \citenamefont {Ren}, \citenamefont {Wang}, \citenamefont {Zhang},
  \citenamefont {H\"{a}nggi},\ and\ \citenamefont {Li}}]{Li2012}%
  \BibitemOpen
  \bibfield  {author} {\bibinfo {author} {\bibfnamefont {N.}~\bibnamefont
  {Li}}, \bibinfo {author} {\bibfnamefont {J.}~\bibnamefont {Ren}}, \bibinfo
  {author} {\bibfnamefont {L.}~\bibnamefont {Wang}}, \bibinfo {author}
  {\bibfnamefont {G.}~\bibnamefont {Zhang}}, \bibinfo {author} {\bibfnamefont
  {P.}~\bibnamefont {H\"{a}nggi}}, \ and\ \bibinfo {author} {\bibfnamefont
  {B.}~\bibnamefont {Li}},\ }\href {\doibase 10.1103/RevModPhys.84.1045}
  {\bibfield  {journal} {\bibinfo  {journal} {Reviews of Modern Physics}\
  }\textbf {\bibinfo {volume} {84}},\ \bibinfo {pages} {1045} (\bibinfo {year}
  {2012})}\BibitemShut {NoStop}%
\bibitem [{\citenamefont {Rieder}\ \emph {et~al.}(1967)\citenamefont {Rieder},
  \citenamefont {Lebowitz},\ and\ \citenamefont {Lieb}}]{Rieder1967}%
  \BibitemOpen
  \bibfield  {author} {\bibinfo {author} {\bibfnamefont {Z.}~\bibnamefont
  {Rieder}}, \bibinfo {author} {\bibfnamefont {J.~L.}\ \bibnamefont
  {Lebowitz}}, \ and\ \bibinfo {author} {\bibfnamefont {E.}~\bibnamefont
  {Lieb}},\ }\href {\doibase 10.1063/1.1705319} {\bibfield  {journal} {\bibinfo
   {journal} {Journal of Mathematical Physics}\ }\textbf {\bibinfo {volume}
  {8}},\ \bibinfo {pages} {1073} (\bibinfo {year} {1967})}\BibitemShut
  {NoStop}%
\bibitem [{\citenamefont {Starr}(1936)}]{Starr1936}%
  \BibitemOpen
  \bibfield  {author} {\bibinfo {author} {\bibfnamefont {C.}~\bibnamefont
  {Starr}},\ }\href {\doibase 10.1063/1.1745338} {\bibfield  {journal}
  {\bibinfo  {journal} {Journal of Applied Physics}\ }\textbf {\bibinfo
  {volume} {7}},\ \bibinfo {pages} {15} (\bibinfo {year} {1936})}\BibitemShut
  {NoStop}%
\bibitem [{\citenamefont {Brattain}(1951)}]{Brattain1951}%
  \BibitemOpen
  \bibfield  {author} {\bibinfo {author} {\bibfnamefont {W.~H.}\ \bibnamefont
  {Brattain}},\ }\href
  {http://scitation.aip.org/content/aip/journal/jap/7/1/10.1063/1.1745338}
  {\bibfield  {journal} {\bibinfo  {journal} {Reviews of Modern Physics}\
  }\textbf {\bibinfo {volume} {23}},\ \bibinfo {pages} {203} (\bibinfo {year}
  {1951})}\BibitemShut {NoStop}%
\bibitem [{\citenamefont {Chang}\ \emph {et~al.}(2006)\citenamefont {Chang},
  \citenamefont {Okawa}, \citenamefont {Majumdar},\ and\ \citenamefont
  {Zettl}}]{Chang2006}%
  \BibitemOpen
  \bibfield  {author} {\bibinfo {author} {\bibfnamefont {C.~W.}\ \bibnamefont
  {Chang}}, \bibinfo {author} {\bibfnamefont {D.}~\bibnamefont {Okawa}},
  \bibinfo {author} {\bibfnamefont {A.}~\bibnamefont {Majumdar}}, \ and\
  \bibinfo {author} {\bibfnamefont {A.}~\bibnamefont {Zettl}},\ }\href
  {\doibase 10.1002/nme.1803} {\bibfield  {journal} {\bibinfo  {journal}
  {Science}\ }\textbf {\bibinfo {volume} {314}},\ \bibinfo {pages} {1121}
  (\bibinfo {year} {2006})}\BibitemShut {NoStop}%
\bibitem [{\citenamefont {Bolsterli}\ \emph {et~al.}(1970)\citenamefont
  {Bolsterli}, \citenamefont {Rich},\ and\ \citenamefont
  {Visscher}}]{Bolsterli1970}%
  \BibitemOpen
  \bibfield  {author} {\bibinfo {author} {\bibfnamefont {M.}~\bibnamefont
  {Bolsterli}}, \bibinfo {author} {\bibfnamefont {M.}~\bibnamefont {Rich}}, \
  and\ \bibinfo {author} {\bibfnamefont {W.~M.}\ \bibnamefont {Visscher}},\
  }\href {http://pra.aps.org/abstract/PRA/v1/i4/p1086\_1} {\bibfield  {journal}
  {\bibinfo  {journal} {Physical Review A}\ }\textbf {\bibinfo {volume} {1}},\
  \bibinfo {pages} {1086} (\bibinfo {year} {1970})}\BibitemShut {NoStop}%
\bibitem [{\citenamefont {Bonetto}\ \emph {et~al.}(2004)\citenamefont
  {Bonetto}, \citenamefont {Lebowitz},\ and\ \citenamefont
  {Lukkarinen}}]{Bonetto2004}%
  \BibitemOpen
  \bibfield  {author} {\bibinfo {author} {\bibfnamefont {F.}~\bibnamefont
  {Bonetto}}, \bibinfo {author} {\bibfnamefont {J.~L.}\ \bibnamefont
  {Lebowitz}}, \ and\ \bibinfo {author} {\bibfnamefont {J.}~\bibnamefont
  {Lukkarinen}},\ }\href@noop {} {\bibfield  {journal} {\bibinfo  {journal}
  {Journal of Statistical Physics}\ }\textbf {\bibinfo {volume} {116}},\
  \bibinfo {pages} {783} (\bibinfo {year} {2004})}\BibitemShut {NoStop}%
\bibitem [{\citenamefont {Bernardin}\ and\ \citenamefont
  {Olla}(2005)}]{Bernardin2005}%
  \BibitemOpen
  \bibfield  {author} {\bibinfo {author} {\bibfnamefont {C.}~\bibnamefont
  {Bernardin}}\ and\ \bibinfo {author} {\bibfnamefont {S.}~\bibnamefont
  {Olla}},\ }\href {\doibase 10.1007/s10955-005-7578-9} {\bibfield  {journal}
  {\bibinfo  {journal} {Journal of Statistical Physics}\ }\textbf {\bibinfo
  {volume} {121}},\ \bibinfo {pages} {271} (\bibinfo {year}
  {2005})}\BibitemShut {NoStop}%
\bibitem [{\citenamefont {Dhar}\ \emph {et~al.}(2011)\citenamefont {Dhar},
  \citenamefont {Venkateshan},\ and\ \citenamefont {Lebowitz}}]{Dhar2011}%
  \BibitemOpen
  \bibfield  {author} {\bibinfo {author} {\bibfnamefont {A.}~\bibnamefont
  {Dhar}}, \bibinfo {author} {\bibfnamefont {K.}~\bibnamefont {Venkateshan}}, \
  and\ \bibinfo {author} {\bibfnamefont {J.~L.}\ \bibnamefont {Lebowitz}},\
  }\href {\doibase 10.1103/PhysRevE.83.021108} {\bibfield  {journal} {\bibinfo
  {journal} {Physical Review E}\ }\textbf {\bibinfo {volume} {83}},\ \bibinfo
  {pages} {021108} (\bibinfo {year} {2011})}\BibitemShut {NoStop}%
\bibitem [{\citenamefont {Landi}\ and\ \citenamefont
  {de~Oliveira}(2014)}]{Landi2014a}%
  \BibitemOpen
  \bibfield  {author} {\bibinfo {author} {\bibfnamefont {G.~T.}\ \bibnamefont
  {Landi}}\ and\ \bibinfo {author} {\bibfnamefont {M.~J.}\ \bibnamefont
  {de~Oliveira}},\ }\href {\doibase 10.1103/PhysRevE.89.022105} {\bibfield
  {journal} {\bibinfo  {journal} {Physical Review E}\ }\textbf {\bibinfo
  {volume} {89}},\ \bibinfo {pages} {022105} (\bibinfo {year}
  {2014})}\BibitemShut {NoStop}%
\bibitem [{\citenamefont {Landi}\ and\ \citenamefont
  {de~Oliveira}(2013)}]{Landi2013a}%
  \BibitemOpen
  \bibfield  {author} {\bibinfo {author} {\bibfnamefont {G.~T.}\ \bibnamefont
  {Landi}}\ and\ \bibinfo {author} {\bibfnamefont {M.~J.}\ \bibnamefont
  {de~Oliveira}},\ }\href {\doibase 10.1103/PhysRevE.87.052126} {\bibfield
  {journal} {\bibinfo  {journal} {Physical Review E}\ }\textbf {\bibinfo
  {volume} {87}},\ \bibinfo {pages} {052126} (\bibinfo {year}
  {2013})}\BibitemShut {NoStop}%
\bibitem [{\citenamefont {Gardiner}\ and\ \citenamefont
  {Zoller}(2004)}]{Gardiner2004}%
  \BibitemOpen
  \bibfield  {author} {\bibinfo {author} {\bibfnamefont {C.}~\bibnamefont
  {Gardiner}}\ and\ \bibinfo {author} {\bibfnamefont {P.}~\bibnamefont
  {Zoller}},\ }\href@noop {} {\emph {\bibinfo {title} {{Quantum noise}}}},\
  \bibinfo {edition} {3rd}\ ed.\ (\bibinfo  {publisher} {Springer},\ \bibinfo
  {year} {2004})\ p.\ \bibinfo {pages} {450}\BibitemShut {NoStop}%
\bibitem [{\citenamefont {Lukkarinen}(2012)}]{Lukkarinen2012}%
  \BibitemOpen
  \bibfield  {author} {\bibinfo {author} {\bibfnamefont {J.}~\bibnamefont
  {Lukkarinen}},\ }\href {\doibase 10.1088/0031-8949/86/05/058507} {\bibfield
  {journal} {\bibinfo  {journal} {Physica Scripta}\ }\textbf {\bibinfo {volume}
  {86}},\ \bibinfo {pages} {058507} (\bibinfo {year} {2012})}\BibitemShut
  {NoStop}%
\bibitem [{\citenamefont {Asadian}\ \emph {et~al.}(2013)\citenamefont
  {Asadian}, \citenamefont {Manzano}, \citenamefont {Tiersch},\ and\
  \citenamefont {Briegel}}]{Asadian2013}%
  \BibitemOpen
  \bibfield  {author} {\bibinfo {author} {\bibfnamefont {A.}~\bibnamefont
  {Asadian}}, \bibinfo {author} {\bibfnamefont {D.}~\bibnamefont {Manzano}},
  \bibinfo {author} {\bibfnamefont {M.}~\bibnamefont {Tiersch}}, \ and\
  \bibinfo {author} {\bibfnamefont {H.~J.}\ \bibnamefont {Briegel}},\ }\href
  {\doibase 10.1103/PhysRevE.87.012109} {\bibfield  {journal} {\bibinfo
  {journal} {Physical Review E}\ }\textbf {\bibinfo {volume} {87}},\ \bibinfo
  {pages} {012109} (\bibinfo {year} {2013})}\BibitemShut {NoStop}%
\bibitem [{\citenamefont {Tom\'{e}}\ and\ \citenamefont
  {de~Oliveira}(2010)}]{tome2010}%
  \BibitemOpen
  \bibfield  {author} {\bibinfo {author} {\bibfnamefont {T.}~\bibnamefont
  {Tom\'{e}}}\ and\ \bibinfo {author} {\bibfnamefont {M.~J.}\ \bibnamefont
  {de~Oliveira}},\ }\href {\doibase 10.1103/PhysRevE.82.021120} {\bibfield
  {journal} {\bibinfo  {journal} {Physical Review E}\ }\textbf {\bibinfo
  {volume} {82}},\ \bibinfo {pages} {021120} (\bibinfo {year}
  {2010})}\BibitemShut {NoStop}%
\bibitem [{\citenamefont {Verlet}(1967)}]{Verlet1967}%
  \BibitemOpen
  \bibfield  {author} {\bibinfo {author} {\bibfnamefont {L.}~\bibnamefont
  {Verlet}},\ }\href {http://prola.aps.org/abstract/PR/v159/i1/p98\_1}
  {\bibfield  {journal} {\bibinfo  {journal} {Physical review}\ }\textbf
  {\bibinfo {volume} {159}},\ \bibinfo {pages} {98} (\bibinfo {year}
  {1967})}\BibitemShut {NoStop}%
\bibitem [{\citenamefont {Gr\o~nbech Jensen}\ and\ \citenamefont
  {Farago}(2013)}]{Gronbech-Jensen2013}%
  \BibitemOpen
  \bibfield  {author} {\bibinfo {author} {\bibfnamefont {N.}~\bibnamefont
  {Gr\o~nbech Jensen}}\ and\ \bibinfo {author} {\bibfnamefont {O.}~\bibnamefont
  {Farago}},\ }\href {\doibase 10.1080/00268976.2012.760055} {\bibfield
  {journal} {\bibinfo  {journal} {Molecular Physics}\ }\textbf {\bibinfo
  {volume} {111}},\ \bibinfo {pages} {983} (\bibinfo {year}
  {2013})}\BibitemShut {NoStop}%
\bibitem [{\citenamefont {Aoki}\ and\ \citenamefont
  {Kusnezov}(2002)}]{Aoki2000a}%
  \BibitemOpen
  \bibfield  {author} {\bibinfo {author} {\bibfnamefont {K.}~\bibnamefont
  {Aoki}}\ and\ \bibinfo {author} {\bibfnamefont {D.}~\bibnamefont
  {Kusnezov}},\ }\href {\doibase 10.1006/aphy.2001.6207} {\bibfield  {journal}
  {\bibinfo  {journal} {Annals of Physics}\ }\textbf {\bibinfo {volume}
  {295}},\ \bibinfo {pages} {50} (\bibinfo {year} {2002})},\ \Eprint
  {http://arxiv.org/abs/0002160} {arXiv:0002160 [hep-ph]} \BibitemShut
  {NoStop}%
\bibitem [{\citenamefont {Aoki}\ \emph {et~al.}(2006)\citenamefont {Aoki},
  \citenamefont {Lukkarinen},\ and\ \citenamefont {Spohn}}]{Aoki2006}%
  \BibitemOpen
  \bibfield  {author} {\bibinfo {author} {\bibfnamefont {K.}~\bibnamefont
  {Aoki}}, \bibinfo {author} {\bibfnamefont {J.}~\bibnamefont {Lukkarinen}}, \
  and\ \bibinfo {author} {\bibfnamefont {H.}~\bibnamefont {Spohn}},\ }\href
  {\doibase 10.1007/s10955-006-9171-2} {\bibfield  {journal} {\bibinfo
  {journal} {Journal of Statistical Physics}\ }\textbf {\bibinfo {volume}
  {124}},\ \bibinfo {pages} {1105} (\bibinfo {year} {2006})},\ \Eprint
  {http://arxiv.org/abs/0602082} {arXiv:0602082 [cond-mat]} \BibitemShut
  {NoStop}%
\end{thebibliography}%
\end{document}